\documentclass[preprint,showpacs,floats,letterpaper,floatfix,,groupedaddress,eqsecnum]{revtex4}
\usepackage{amssymb,amsmath}
\usepackage[dvips]{graphicx}

\usepackage{dcolumn,epsfig}

\begin{document}

\title{ Study of Instability of Liquid Jets Under Gravity}
\author{Wellstandfree K Bani, and Mangal C. Mahato}
\email{mangal@nehu.ac.in}
\affiliation{Department of Physics, North-Eastern Hill University,
Shillong-793022, India}

\begin{abstract}
Breakup of water jets under gravity is a common-place phenomenon. The role of 
surface tension in the instability of water jets was recognized by Rayleigh 
and the theory propounded goes by the name of Plateau-Rayleigh theory. The 
necks and bulges down along the jet-length that are created by perturbation 
waves of wavelengths larger than a certain value keep growing with time and 
ultimately cause the jet to breakup into drops. The effect of perturbation 
waves have been investigated experimentally and found to confirm the 
essentials of the theory. However, there is no unanimity about the origin of 
these perturbation waves. Recently, the idea of recoil capillary waves as an 
important source of the perturbation waves has been emphasized. The recoil of 
the end point of the remaining continuous jet at its breakup point is 
considered to travel upward as a recoil capillary wave which gets reflected at 
the mouth of the nozzle from which the jet originates. The reflected capillary 
wave travels along the jet downward with its Doppler shifted wavelength as a 
perturbation wave. We set up an experiment to directly verify the existence 
and effect of the recoil capillary waves and present some preliminary results 
of our experiment.
\end{abstract}

\vspace{0.5cm}
\date{\today}

\maketitle

\section{Introduction}

A continuous water jet falling gently downward through a tap and its 
subsequent breakup into discontinuous drops is an everyday-witnessed common 
phenomenon. There have been many attempts theoretically as well as 
experimentally to understand the physical mechanism behind the breaking up of
jets into drops instead of continuing to keep reducing its cross-sectional area 
forever. However, the complete understanding of this macroscopic phenomenon 
has eluded so far.

The scientific study of this phenomenon has a long  history. Savart (1833) 
experimentally observed the instability of liquid jets that clearly indicated 
that the liquid column (jet) undergoes changes before breaking up into disjoint 
droplets. Savart’s experimental observations suggested that the liquid column 
develops necks and bulges in the form of waves which ultimately leads to the 
formation of disjoint droplets out of the continuous column. This problem of 
instability of liquid jets was taken up by Plateau and then by Rayleigh (1879) 
and developed a theory which came to be known as Plateau-Rayleigh theory. 
Rayleigh\cite{Ray1,Ray2} showed that perturbations of wavelengths larger than 
a certain value (larger than the jet diameter) grow rapidly with time which 
ultimately make the column unstable (Rayleigh instability) against formation 
of droplets based on surface energy considerations of inviscid liquids. 
Chandrasekhar\cite{Chandra} later extended the theory to viscous liquids. Many 
experimental investigations have been conducted to examine the validity of 
Plateau-Rayleigh theory\cite{Goedde}. The investigations on the instability of
water jet and related phenomena have been extensively 
reviewed\cite{Eggers1,Eggers2,Lin}. The study continues to be of current 
interest\cite{Rajendran}.

The Plateau-Rayleigh theory, describes how the response of perturbations of 
certain wavelengths keep growing with time whereas those with lower 
wavelengths decay with time. It is the fastest growing perturbation that
utimately breaks the jet. However, the theory is silent about the origin of 
the perturbations. The experiment of Goedde and Yuen, for example, applied 
external perturbations to study the length of the liquid jet (measured from the 
root of the jet) before it breaks up. However, even if no external 
perturbations are applied, one obtains a finite deterministic length of the 
liquid jet. In this case, obviously, the perturbation must have an origin at 
the root of the jet at the mouth of the nozzle, through which the liquid flows
out to form the jet. Naturally, the perturbations will have amplitudes of 
random sizes. As a consequence, the length of the jet should have random 
values. However, on the average, the length turns out to be a 
deterministically fixed number depending on the conditions of the experiment. 
In some recent works, in order to overcome this difficulty of fixing the 
origin of the perturbations, Umemura and co-workers\cite{Umemura1,Umemura2}
emphasized the idea that the perturbations get generated and sustained 
self-consistently. This is based on the observed fact that soon after the jet 
breaks up (at one of the points on the lowermost neck) the tip of the 
remaining column contracts to make its shape round once again to minimize the 
surface energy. The tip contraction gives rise to an upstream propagating 
capillary wave which upon reflection at the mouth of the nozzle moves 
downstream with Doppler modified wavelengths. Some of these waves with the 
right wavelength (as discussed by Rayleigh) cause the liquid column to break 
producing another contraction of the tip of the column and the process repeats 
once again. We set up and conduct an experiment to verify the existence and 
effect of recoil capillary waves on the length of continuous water jet. This 
we accomplish by damping the recoil capillary wave using an earlier used 
method\cite{Mahato}.
 
\section{Experimental set up and experiment}

\begin{figure}[htp]
\centering
\includegraphics[width=10cm,height=7cm]{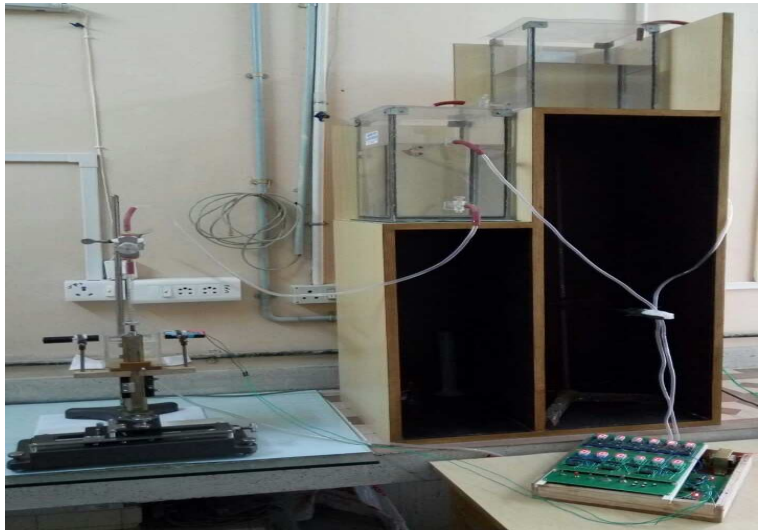}
\caption{Photograph of the experimental set up.}
\end{figure}

\begin{figure}[htp]
\centering
\includegraphics[width=10cm,height=7cm]{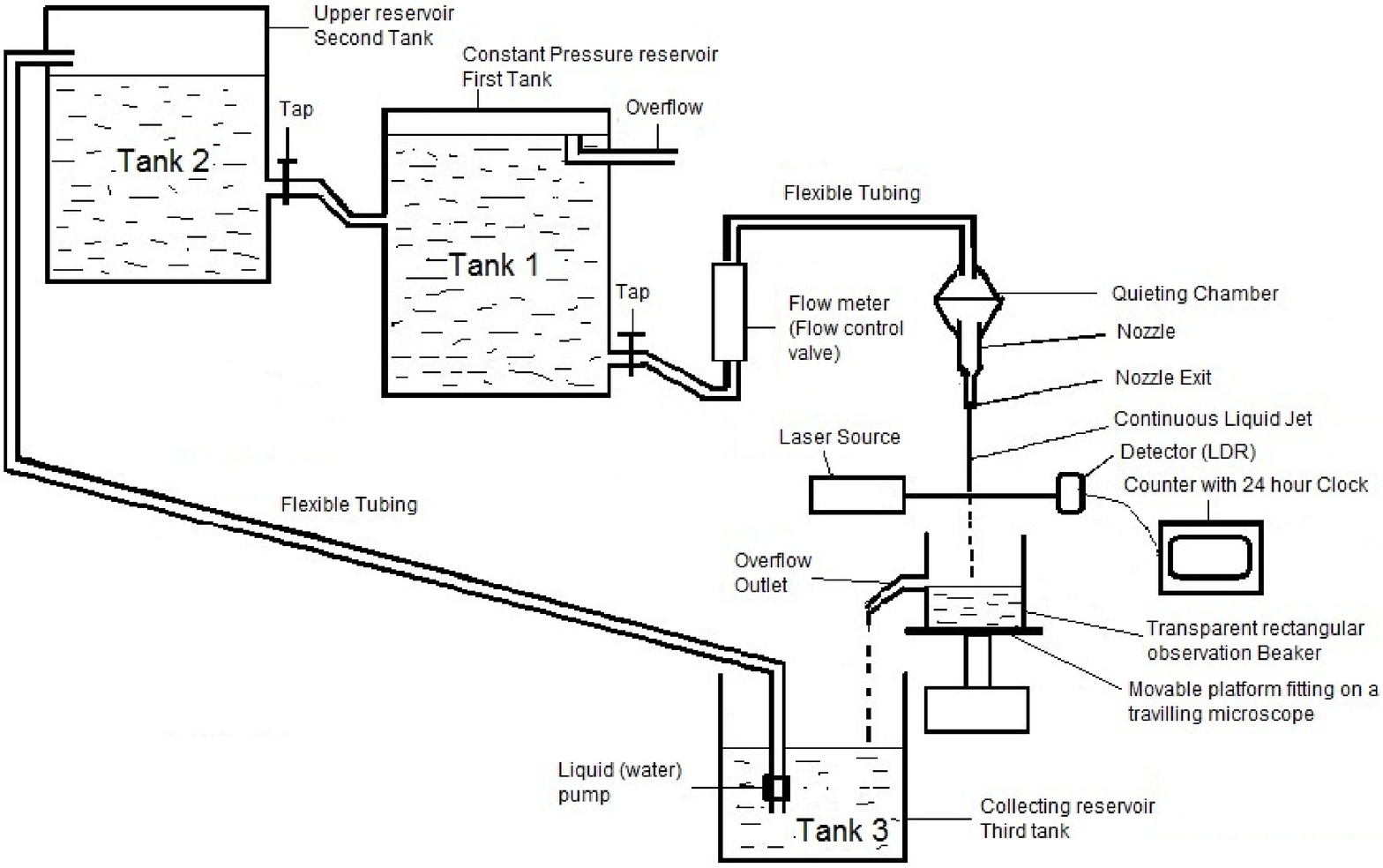}
\caption{The diagramatic sketch of the experimental set up.}
\end{figure}

The experimental set up (Figs.1,2) consists of two water tanks, the first one 
(Tank 1) is placed at a lower height to maintain the constant level of water 
(i.e., to keep same pressure throughout the experiment) and the second one 
(Tank 2) is placed at a higher height to supply water at a constant rate to 
the Tank 2. Both of these tanks are fitted with one tap each. The tap of the 
Tank 2 is connected to the inlet of Tank 1 by a rubber pipe. A glass nozzle 
is connected to a tap of the Tank 1 by a rubber pipe through a flow control 
valve, which permits us to control the flow rate of water. The excess water in 
the Tank 1 flows out through the level outlet to the lower reservoir tank 
(Tank 3) (this tank is kept at the bottom level). The water from the Tank 3 is 
pumped up to the Tank 2 by using a water pump. A transparent rectangular 
beaker with a level outlet on one of its vertical sides is placed vertically 
below the nozzle so that the water jet falls directly on the water kept in the 
beaker. The water level in the beaker is maintained fixed by letting the 
excess water flow out through the level outlet of the beaker. The beaker is 
placed on a horizontal platform fitted to a travelling microscope so that the 
beaker can be smoothly moved vertically up and down and its position measured. 
A  laser-pointer-and-detector arrangement is also fitted to the platform so 
that the horizontal laser beam incident normally on the vertical surface of the beaker and passes through the path of the water jet and then through the 
opposite surface of the beaker before it is collected by the detector. The 
laser-detector arrangement can be moved vertically and fixed as desired and 
the vertical position of the horizontal laser beam measured. A six digit 
counter mounted with a 24-hour clock is connected to the detector to count the 
number of drops and also to pin-point the first initial breaking point of the 
jet. Our experimental set up is similar in essentials to that of Goedde and 
Yuen (compare Fig.1 of Ref.\cite{Goedde}). The ambient condition of 
temperature and humidity is controlled and allowed to settle down at desired 
values before the experiment is conducted and the condition is kept fixed 
throughout the duration of experiment.

The water is issued (jetted) vertically downward through a long glass nozzle 
(of length larger than about ten times its internal diameter, $d$) in to the water in the beaker
through the stagnant ambient air atmosphere . As 
long as the jet remains continuous the detector remains quiescent. However, a 
breakup into a drop (disjoint of the jet) is detected as a count. The vertical distance between the 
nozzle exit and the laser beam gives the breakup length. Initially, the laser 
beam is made to face the continuous jet by moving the platform up and then the 
platform is gradually lowered so that the beam position distance from the 
nozzle exit increases. At every position of the laser-detector arrangement the 
counts are recorded for two minutes for several times and their average 
calculated to obtain the average number of drop-counts per second. The 
platform is slowly and gradually lowered by small steps and at each position 
the experiment is repeated to obtain the average count rate. Naturally, the 
count rate begins from zero, reaches a threshold at which the rate just begins 
to show nonzero value and then gradually keeps increasing as the platform is 
slowly lowered in small steps. The process continued till the rate reaches a 
maximum (saturation) value. Throughout the above process the water flow rate 
is kept fixed. The same process is then repeated for several values of flow 
rates. Note that after each change of flow rate the flow and the jet is 
allowed to become steady before the measurement process is begun.

For our purpose, we perform two distinct kinds (sets) of experiments. In one 
we let the laser beam pass grazing the water level in the beaker (just about 
0.2 cm above the water surface on the beaker). In the other the beam is kept 
at a height of about 1.5 cm above the water surface. In the first set of 
experiments, if the point of separation between a drop and the tip of the 
remaining jet is such that before the moving tip recoils it just touches the 
water surface, the process of jet-tip recoil is prevented and thus the recoil 
capillary wave propagating up the jet length is damped whereas in the other 
set of experiments this situation never occurs.

The experiment is performed in an enclosure where the air current is minimized 
and, as stated earlier, the temperature and humidity are kept fixed for all 
sets of measurements. Thus, the enclosure inside the lab is made sure to be 
stagnant and the external disturbance on the jet is minimized.

\section{Experimental results}

\begin{figure}[htp]
\centering
\includegraphics[width=10cm,height=7cm]{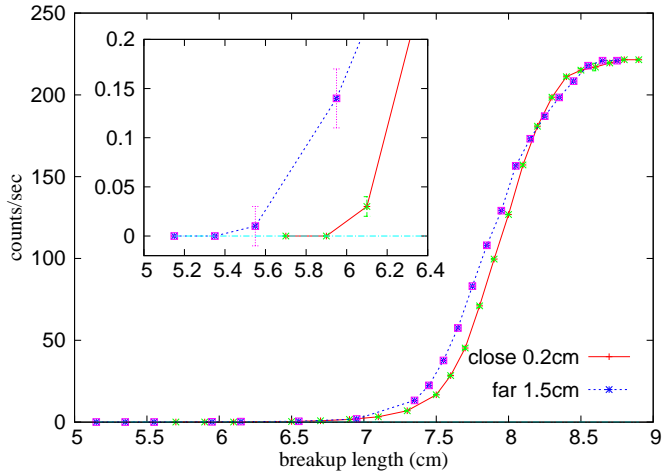}
\caption{Count rate (s$^{-1}$) as a function of distance of the laser beam from
the nozzle exit at the flow rate of 55.5 cc/min and $d=1.71$ mm for the two 
sets of experiments (laser beam close to the water surface and far from the
surface. The inset shows a magnified graph to show the first (jet) breakup 
point.}
\end{figure}

\begin{figure}[htp]
\centering
\includegraphics[width=10cm,height=7cm]{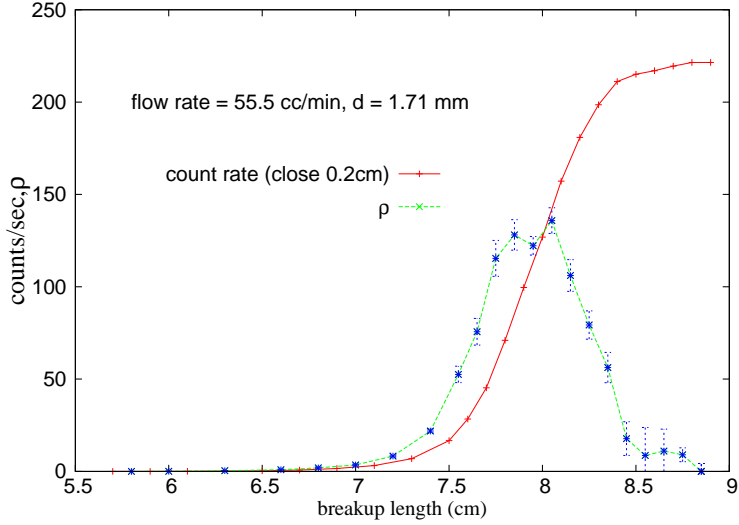}
\caption{Count rate (s$^{-1}$)  as a function of distance of the laser beam 
from the nozzle exit (same as Fig.3 but only for the set close to the surface).
It also shows the corresponding unnormalized breakup length distribution.}
\end{figure}

Figure 3 shows the average number of counts (drops) per second, for a water 
flow rate of 55.5 cc/min and the nozzle inner diameter of 1.71 mm, as the 
vertical distance of the laser beam from the lower tip of the nozzle is 
increased. Each point on the graph corresponds to an average of three set of 
counts for a duration of two minutes each. The counts begin from zero 
(indicating continuous jet) to a saturation value (indicating the finality of 
the breakup process). From this data we calculate the distribution, $\rho$, 
of breakup lengths. The probabilty density of breakup lengths is roughly 
calculated as the derivative of the count rate with respect to breakup length, 
Fig. 4. In Fig.4 the distribution $\rho$ is not normalized. The normalized 
distribution $\rho_n$ will be obtained as $\rho_n=\rho /100$. The distribution 
provides information about the mean value of breakup lengths. The mean values 
are plotted in Fig.5 for the same nozzle of inner diameter 1.71 mm but for different 
flow rates for the two sets of experiments (with and without damping). The two 
sets of mean values lie in a band. The approximate equality of the two mean 
values is understandable, however, because once the water level in the 
observation beaker is “well below” the breakup point, it is as good as being 
far below and equivalent to a point of the set without damping.

The inset of Fig.3 gives a magnified diagram of the curves of Fig. 3 in the
lower range of the distance of laser beam from the nozzle exit. From this
graph one can obtain the first (initial) breakup lengths in the two cases
(with and without damping) and are plotted in Fig. 5 as a function of water
flow rates. The figure clearly shows that the first breakup length with the
damped recoil capillary wave is larger than that without the damping for
all values of flow rates measured. This indicates that the recoil capillary
waves do exist and they aid in making the water jet unstable against breakup.

The mean breakup lengths plotted in Fig.5, though not exactly identical to 
the earlier reported results (for example, p. 104 of Ref.\cite{Lin}), the 
qualitative features are very similar, showing various regimes of jet breakup.
However, information on the first breakup length is entirely new. It shows that 
the first breakup lengths peak at a smaller Reynolds number (mean Re=1079.3) 
and Weber number (We=7.54) than in case of the mean breakup lengths (Re=1392.6,
mean We=12.51). Our measurements are done at the temperature of 
(25$\pm$.5)$^{\circ}$C and at relative humidity of 80$\pm$3\%. However, in 
order to calculate Re and We, we have used the tabulated values of surface 
tension of water $\sigma=72\times10^{-3}$Nm$^{-1}$ and coefficient of viscosity 
$\mu=8.9\times 10^{-4}$ kg m$^{-1}$s$^{-1}$. 

\begin{figure}[htp]
\centering
\includegraphics[width=10cm,height=7cm]{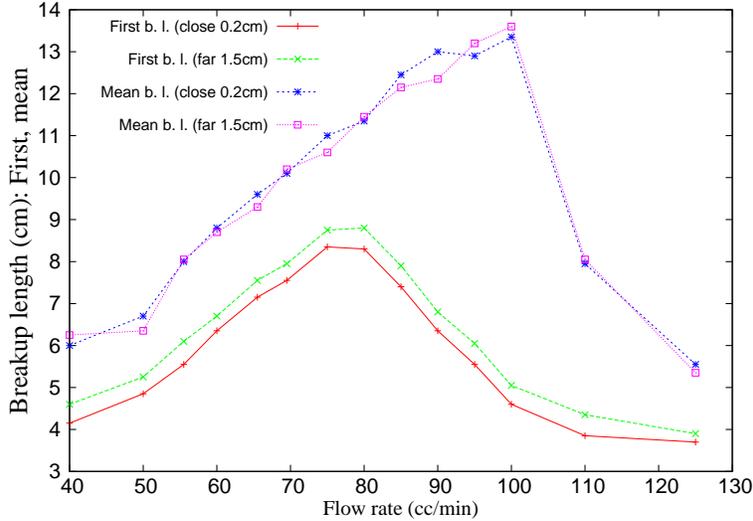}
\caption{First (lower set of two curves) and mean (upper set of two curves) 
breakup lengths (b.l.) as a function of water flow rate through the same nozzle
of internal diameter, d=1.71 mm.}
\end{figure}

The inset of Fig.3 gives a magnified diagram of the curves of Fig. 3 in the
lower range of the distance of laser beam from the nozzle exit. From this 
graph one can obtain the first (initial) breakup lengths in the two cases 
(with and without damping) and are plotted in Fig. 5 as a function of water 
flow rates. The figure clearly shows that the first breakup length with the 
damped recoil capillary wave is larger than that without the damping for 
all values of flow rates measured. This indicates that the recoil capillary 
waves do exist and they aid in making the water jet unstable against breakup.

\section{Conclusion}
We have performed a very simple experiment the results of which may be 
considered to directly give a proof of the existence of recoil capillary 
waves and their effect on the jet breakup length. Though the experiment is not 
so precise and sophisticated the qualitative features shown are unmistakable. 
A high speed photographic measurement may help in arriving at a result with 
more confidence. We hope to obtain a conclusive result once we have the 
required facility.

\end{document}